# Topological materials or structures: Origin of higher-order topological states

Shengjie Zheng[1,2,3], Guiju Duan[1,2,3], Jianting Liu[1,2], Baizhan Xia[1,2 ✉]

[1]State Key Laboratory of Advanced Design and Manufacturing for Vehicle Body, Hunan University, Changsha, Hunan 410082, China

[2]College of Mechanical and Vehicle Engineering, Hunan University, Changsha, Hunan 410082, China

[3]These authors contributed equally: Shengjie Zheng, Guiju Duan.

✉Corresponding author. Email: xiabz2013@hnu.edu.cn.

**Abstract:** Higher-order topological states (HOTS) have been extensively investigated in classical wave systems. They do not exist in the band gaps of infinite materials, while exhibit as the in-gap localized modes once the infinite materials are truncated to be the finite structures. Here, we will experimentally reveal the origin of HOTSs in acoustic systems. We design the hollow acoustic structures exclusively composed of hinge and corner resonators. We present the experimental proof that, despite the lack of surfaces and bulks, the hollow acoustic structures can still support the topologically protected hinge and corner states, indicating that the local configurations of boundaries are the sources for the generation of HOTSs. We then get the composite structures by assembling the 2D and 3D topological hollow acoustic structures, and experimentally observe the robust HOTSs in them. Our results provide a fundamental perspective on HOTSs in periodic structures, and we foresee that these findings will pave the way toward designing new topological devices for energy recovering, information processing, non-destructive testing and acoustic sensing.

HOTSs, which express as scattering-free hinge and corner states surrounding insulating bulks, have been theoretically forecasted[1-4], and numerically and experimentally realized in various of physics, such as electronics[5, 6], microwaves[7, 8], photonics[9-14], electrics[15, 16], acoustics[17-25], and mechanics[26, 27]. Recently, HOTSs have been extended to crystalline lattices with defects[28-30]. HOTSs are featured by the bulk polarizations and topological invariances[2, 25, 31]. It is well known that HOTSs cannot be existence when the materials are infinite ones without boundaries. If the infinite materials are truncated to be the finite structures, HOTSs can be generated in the structures with special boundaries conditions. Thus, we have reason to believe that the local configurations of boundaries of materials play intrinsic roles on HOTSs.

The acoustic structures with the high designability provide the flexible platforms to research the origin of HOTSs. We firstly design the 2D and 3D topological hollow acoustic structures exclusively composed of hinge and corner resonators. Despite the absence of actual bulks and surfaces, we experimentally feature the topologically protected hinge and corner states via pump-probe measurements. The HOTSs in topological hollow acoustic structures are exactly the same as those in traditional topological acoustic structures with bulks and surfaces, which indicates that the local configurations of boundaries are the sources for the generation of HOTSs. Then, we get the composite structures by assembling the 2D and 3D hollow acoustic structures, and present the experimental proof that, the higher-order topological hinge and corner states can be excited in composite structures with complex boundaries. We unveil that these findings with simulations and experiments will open the new way toward the study of the HOTSs in various topological systems.

## Lattice models

To demonstrate the idea, let us consider the simple 2D and 3D lattice models in Figs. 1a and 1b. The light blue (light yellow) colored bond represents a negative (positive) hopping. The thick (thin) bond has a strong (weak) hopping amplitude. There are two types of unit cells. The first one with strong intra-unit-cell couplings is the trivial unit cell and the other one with weak intra-unit-cell couplings is the nontrivial unit cell [2, 31]. Both unit cells have the same band structures. For a material with infinite atoms, its topological properties cannot be described by deliberately defining trivial and nontrivial unit cells, as they are essentially same and any one of them can form the material. The amazing evolution is generated when the material is truncated to a finite structure with well-defined boundaries. The lattice structures are shown in Figs. 1c and 1d, which are consisted of the unit cells with strong intra-unit-cell couplings and weak intra-unit-cell couplings, respectively. When the intra-unit-cell couplings are seriously stronger than the inter-unit-cell couplings, the lattice structures are trivial ones without any HOTS, as shown in Fig. 1e. The lattice nodes with stronger couplings can be considered as a polymer. Thus, the lattice structures in Fig. 1c are consisted of tetramers (for the 2D one) and octamers (for the 3D one). The attributes of the lattice nodes of structures from the bulks to the boundaries are the same and there will be no localized states within the band gaps of the unit cell. When the intra-unit-cell couplings are seriously weaker than the inter-unit-cell couplings, the lattice structures are nontrivial ones with HOTS, as shown in Fig. 1f. The 2D structure in Fig. 1d can be redivided to the monomers (at the corners), dimers (along the hinges) and tetramers (on the surfaces), according to the coupling strengths among the lattice nodes. Similarly, the 3D structure in Fig. 1d can be redivided to the monomers (at the corners), dimers (along the hinges), tetramers (on the surfaces) and octamers (in the bulk). The attributes of lattice nodes at the boundaries (corners, hinges and surfaces) are different from those in the bulks (seeing Fig. S1 in Supplemental Material). Thus, the new localized states, exhibiting as the topologically protected corner, hinge and surface states, will appear in the band gaps of the unit cell in the bulk. The trivial and nontrivial structures can be transformed into each other, and evolved into a hybrid structure with partial HOTS, by modulating the boundaries of structures (seeing Fig. S2 in Supplemental Material).

## 2D hollow acoustic structure without surface

We start with a conventional 2D acoustic structure building on the square lattice with quantized quadrupole phases[2, 31]. The cavities represent the artificial atoms. The light blue (light yellow) tubes represent hopping terms with negative (positive) signs. The negative and positive hoppings are realized by connecting the thin waveguides to different sides of each resonance's dipole nodal line, seeing Fig. S3 in Supplemental Material[32]. There are four resonators in each unit-cell. We use the thin (thick) tubes for the intra-unit-cell (inter-unit-cell) couplings (Fig. 2a). Thus, the intra-unit-cell couplings ($t_1$) are weaker than the inter-unit-cell couplings ($t_2$). The topological phase of this lattice has been defined by its bulk topological invariants which are the polarizations along the $x$ and $y$ directions [2, 25, 31]. We calculate its acoustic eigenvalue spectrum using full-wave finite-element simulations. Results presented in Fig. 2d show that the numbers of the hinge and corner states are 16 and 4 which are respectively consistent with the numbers of resonators along the 4 hinges and at the 4 corners of the acoustic structure. Besides, there are 16 surface states, being consistent with the number of resonators on the surface of the acoustic structure.

To yield a 2D topological insulator without surface states, we delete the resonators on the surface of the acoustic structure (Fig. 2b), leaving only 20 resonators along the boundaries of the acoustic structure (Fig. 2c). This 2D acoustic structure without the surface is called as the 2D hollow acoustic structure here. Comparing its numerically calculated eigenvalue spectrum (Fig. 2f) with that of the conventional acoustic structure (Fig. 2d), the notable differences is that the surface states vanish as expected, due to the absent of the surface. The 16 hinge states and the 4 corner states almost keep the same eigenfrequencies as before, seeing Fig. S4 in Supplemental Material.

In order to reveal the generation of HOTSs in essence, we deeply investigated the wave characteristics of resonators in the 2D acoustic structure. As shown in Fig. 2a, this 2D acoustic structure can be divided into three components according to the strengths among resonators. The first one is the tetramer including four resonators. The second one is the dimer including two resonators. The third one is the monomer including a resonator. As the couplings of these components with adjacent ones are weak, they can be approximately considered as the independent structures. For the tetramer (shown in Fig. 2g), there are 4 eigenstates whose eigenfrequencies are consistent with those of the surface states shown in Figs. 2d and 2e. Thus, in essence, the 4 tetramers of the acoustic structure support the 16 (4 by 4) surface states. The similar results can be obtained in the dimmers, as shown in Fig. 2h. Namely, the 8 dimmers at the edge boundaries yield the 16 (2 by 8) edge states. The monomer is an independent resonator. It has an eigenstate at 3457.4Hz, as shown in Fig. 2i. Therefore, the 4 corner states are inherently generated from the 4 monomers. These results indicate that the HOTSs in the 2D acoustic structures with quantized multipole phases are essentially depended on the local configurations of the boundaries. If the tetramers are deleted, the dimers and the monomers along the boundaries can still support the HOTSs.

## 3D hollow acoustic structure without bulk and surfaces

We then construct a conventional 3D acoustic structure in the cube lattice with quantized octupole phases[2, 31]. The light blue (light yellow) tubes represent hopping terms with negative (positive) signs. There are eight resonators in each unit-cell. Using the thin (thick) tubes for the inter-unit-cell (intra-unit-cell) couplings (Fig. 3a), this acoustic structure is also a topological one with higher-order topological phases[2, 25, 31]. Its simulated acoustic eigenvalue spectrum is presented in Fig. 3d. It shows that there are 64 bulk states, 96 surface states, 48 hinge states and 8 corner states which are respectively consistent with the number of resonators in the bulk, on the 6 surfaces, along the 12 hinges and at the 8 corners of the 3D acoustic structure.

To yield a 3D topological insulator with only hinge and corner states, we remove the resonators in the bulk and on the 6 surfaces (Fig. 3b), leaving only 48 resonators along the hinges and 8 resonators at the corners (Fig. 3c). This 3D acoustic structure without the bulk and surfaces is called as the 3D hollow acoustic structure here. Its simulated acoustic eigenvalue spectrum presented in Fig. 3f shows that the bulk states and the surface states vanish as expected, due to the absent of the bulk and surfaces. Interestingly, there are no fundamental change for all 48 hinge states and 8 corner states, including their eigenfrequencies and their eigenmodes, when compared with the calculated eigenvalue spectrum (Fig. 3d) of the conventional 3D acoustic structure, seeing Fig. S5 in Supplemental Material. If only the 64 resonators in bulk are deleted, an eigenvalue spectrum which consists of 96 surface states, 48 hinge states and 8 corner states will be produced, seeing Fig. S6 in Supplemental Material.

This 3D acoustic structure shown in Fig. 3a can be divided into four components, namely octamers, tetramers, dimers and monomers whose eigenstates are shown Fig. S7 in in Supplemental Material. The HOTSs in the 3D acoustic structures are inherently depended on these components. In detail, the 48 hinge states are generated from the 24 dimers distributing along the 12 hinges and the 8 corner states are generated from the 8 monomers distributing on the 8 corners. The 8 octamers in the bulk yield the 64 (8 by 8) bulk states and the 24 tetramers on 6 surfaces yield the 96 (24 by 4) surface states. Therefore, the HOTSs in the 3D acoustic structures with quantized multipole phases are also depended on the local configurations of the boundaries. Deleting the octamers in the bulk and the dimers on the surfaces, the dimers and the monomers of the 3D acoustic structures can still support the higher-order topological hinge and corner states.

## Experimental realization

The hollow acoustic structure is a new type of topological insulator without the bulk and surfaces. Compared with the conventional acoustic structures, the new developed hollow acoustic structures which vacate their interior spaces will greatly contribute to the integrated design of multifunctional devices. The designed hollow acoustic structures are fabricated by the 3D printing technique (see Figs. 4a and 4b). The manufacturing details are presented in methods in Supplemental Material. We detect the eigenstates via acoustic pump-probe measurements, by placing a small speaker at one side of the cavity and a tiny microphone at the other side of the same cavity (see details in methods in Supplemental Material). In such a pump-probe measurement, the measured response spectra approximately give the local densities of states for the acoustic phonons in the measured cavities. In Fig. 4c, we show the hinge response spectra of the 2D (grey line) and 3D (red line) hollow acoustic structures. These hinge response spectra have two peaks around 3345 Hz and 3571 Hz which are coincident with the eigenfrequencies of hinge states in Figs. 2f and 3f. For the corner response spectra of the 2D (blue line) and 3D (green line) hollow acoustic structures, there is only one peak at around the frequency of 3464 Hz (in Fig. 4c), which is consistent with the simulated eigenfrequencies of corner states in Figs. 2f and 3f.

In addition, we measured the acoustic distributions (specifically, acoustic pressure amplitudes) of the corner state (3464 Hz) and the hinge state (3571 Hz) of the hollow acoustic structures. Cavities at the top surface are excited and tested one by one (see details in methods in Supplemental Material). As shown in Fig. 4d, for the hinge (corner) states, the acoustic energy is strongly localized at the hinge (corner) cavities. Then, we further measured the acoustic pressure phases at the hinge and corner cavities. Due to the dipole nature of the hinge and corner eigenmodes (in detail, the hinge and corner eigenmodes examined here are associated with the lowest orbitals along the major axis of the cuboid cavity), the acoustic phases at the left and right halves of the cavity differ by $\pi$, as shown in Fig. 4e. These results are perfectly matched with the topological phenomenon measured in the conventional topological acoustic structures (presented in the Fig. S8 in Supplemental Material), experimentally proving that the HOTSs of the topological hollow acoustic structures are inherently the same as those of the conventional topological acoustic structures.

Finally, we design a composite structure with complex boundaries, by assembling the 2D and 3D hollow acoustic structures, as shown in Fig. 5a. The 2D and 3D hollow acoustic structures share a hinge. The acoustic pressure amplitude distributions of the topological hinge and corner states are presented in Figs. 5b and 5c, by exciting and testing cavities on the top surface one by one. When the hinge (corner) state is excited, the acoustic energy is strongly localized at the hinge (corner) cavities of the 2D and 3D hollow acoustic structures. These results indicate that the topologically protected hinge and corner states can be realized in the composite structure with complex boundaries. The similar topological characteristics can be also found in the composite structures composed of the conventional 2D and 3D acoustic structures, shown in Fig. S9 in Supplemental Material. Other composite structures composed of the 2D and 3D acoustic structures with slipping dislocations can also stimulate the topological hinge and corner states (shown in Figs. 5d-f and Fig. S10 in Supplemental Material) which further verify the robustness of HOTSs.

## Discussion

In summary, we have designed the topological insulators exclusively composed of hinge and corner resonators, and experimentally proved that despite the lack of surfaces and bulks, the 2D and 3D hollow acoustic structures can still support the HOTSs, verifying that the HOTSs are depended on the local configurations of the boundaries of structures. The higher-order topological structures without the bulk and surfaces will greatly save the physical space, being of great significance to the integrated design of multifunctional devices. Assembling the hollow acoustic structures with different dimensions as a whole, we presented the experimental proof of the HOTSs in composite structures with complex boundaries. Our fundamental perspective on the HOTSs of periodic structures have far-reaching scientific implications for designing complex acoustic devices. When scaled to photon and quantum, our findings would greatly improve the design freedom of photonic devices and quantum devices for topological lasing, electromagnetic communication and quantum computing.

## References

1. Song, Z., Fang, Z. & Fang, C. (d − 2)-dimensional edge states of rotation symmetry protected 437 topological states. *Phys. Rev. Lett.* **119**, 246402 (2017).
2. Benalcazar, W. A., Bernevig, B. A. & Hughes, T. L. Quantized electric multipole insulators. *Science* **357**, 61–66 (2017).
3. Schindler, F. et al. Higher-order topological insulators. *Sci. Adv.* **4**, eaat0346 (2018).
4. Ezawa, M. Higher-order topological insulators and semimetals on the breathing kagome and pyrochlore lattices. *Phys. Rev. Lett.* **120**, 026801 (2018).
5. Tiwari, A., Li, M.-H., Bernevig, B. A., Neupert, T. & Parameswaran, S. A. Unhinging the surfaces of higherorder topological insulators and superconductors. *Phys. Rev. Lett.* **124**, 046801 (2020).
6. Kempkes, S. et al. Robust zero-energy modes in an electronic higher-order topological insulator. *Nat. Mater.* **18**, 1292–1297 (2019).

7. Peterson, C. W., Benalcazar, W. A., Hughes, T. L. & Bahl, G. A quantized microwave quadrupole insulator with topologically protected corner states. *Nature* **555**, 346–350 (2018).
8. Peterson, C. W., Li, T., Benalcazar, W. A., Hughes, T. L. & Bahl, G. A fractional corner anomaly reveals higherorder topology. *Science* **368**, 1114–1118 (2020).
9. Xie, B. et al. Higher-order quantum spin Hall effect in a photonic crystal. *Nat. Commun.* **11**, 3768 (2020).
10. El Hassan, A. et al. Corner states of light in photonic waveguides. *Nat. Photonics* **13**, 697–700 (2019).
11. Li, M. et al. Higher-order topological states in photonic kagome crystals with long-range interactions. *Nat. Photonics* **14**, 89–94 (2020).
12. Noh, J. et al. Topological protection of photonic mid-gap defect modes. *Nat. Photonics* **12**, 408–415 (2018).
13. Cerjan, A., Jürgensen, M., Benalcazar, W. A., Mukherjee, S. & Rechtsman, M. C. Observation of a higher-order topological bound state in the continuum. *Phys. Rev. Lett.* **125**, 213901 (2020).
14. Xie, B.-Y. et al. Visualization of higher-order topological insulating phases in two-dimensional dielectric photonic crystals. *Phys. Rev. Lett.* **122**, 233903 (2019).
15. Imhof, S. et al. Topolectrical-circuit realization of topological corner modes. *Nat. Phys.* **14**, 925–929 (2018).
16. Bao, J. et al. Topoelectrical circuit octupole insulator with topologically protected corner states. *Phys. Rev. B* **100**, 201406 (2019).
17. Zhang, X. et al. Dimensional hierarchy of higher-order topology in three-dimensional sonic crystals. *Nat. Commun.* **10**, 5331 (2019).
18. Ni, X., Weiner, M., Alu, A. & Khanikaev, A. B. Observation of higher-order topological acoustic states protected by generalized chiral symmetry. *Nat. Mater.* **18**, 113–120 (2019).
19. Xue, H., Yang, Y., Gao, F., Chong, Y. & Zhang, B. Acoustic higher-order topological insulator on a Kagome lattice. *Nat. Mater.* **18**, 108–112 (2019).
20. Du, J., Li, T., Fan, X., Zhang, Q. & Qiu, C. Acoustic realization of surface-obstructed topological insulators. *Phys. Rev. Lett.* **128**, 224301 (2022).
21. Wei, Q. et al. 3D hinge transport in acoustic higher-order topological insulators. *Phys. Rev. Lett.* **127**, 255501 (2021).
22. Wu, X. et al. Topological corner modes induced by dirac vortices in arbitrary geometry. *Phys. Rev. Lett.* **126**, 226802 (2021).
23. Yang, Y. et al. Hybrid-order topological insulators in a phononic crystal. *Phys. Rev. Lett.* **126**, 156801 (2021).
24. Ni, X., Li, M., Weiner, M., Alù, A. & Khanikaev, A. B. Demonstration of a quantized acoustic octupole topological insulator. *Nat. Commun.* **11**, 2108 (2020).
25. Xue, H. et al. Observation of an acoustic octupole topological insulator. *Nat. Commun.* **11**, 2442 (2020).
26. Serra-Garcia, M. et al. Observation of a phononic quadrupole topological insulator. *Nature* **555**, 342–345 (2018).
27. Fan, H., Xia, B., Tong, L., Zheng, S. & Yu, D. Elastic higher-order topological insulator with topologically protected corner states. *Phys. Rev. Lett.* **122**, 204301 (2019).
28. Liu, Y. et al. Bulk-disclination correspondence in topological crystalline insulators. *Nature* **589**, 381–385 (2021).
29. Peterson, C. W., Li, T., Jiang, W., Hughes, T. L. & Bahl, G. Trapped fractional charges at bulk defects in topological insulators. *Nature* **589**, 376–380 (2021).

30. Lin, Z.-K. et al. Topological Wannier cycles induced by sub-unit-cell artificial gauge flux in a sonic crystal. *Nat. Mater.* **21**, 430–437 (2022).
31. Benalcazar, W. A., Bernevig, B. A. & Hughes, T. L. Electric multipole moments, topological multipole moment pumping, and chiral hinge states in crystalline insulators. *Phys. Rev. B* **96**, 245115 (2017).

## Acknowledgements

This work was supported by the Natural Science Foundation of Hunan Province for Distinguished Young Scholars (Grant No. 2022JJ10003) and the National Natural Science Foundation of China (Grant Nos. 12072108, 51621004).

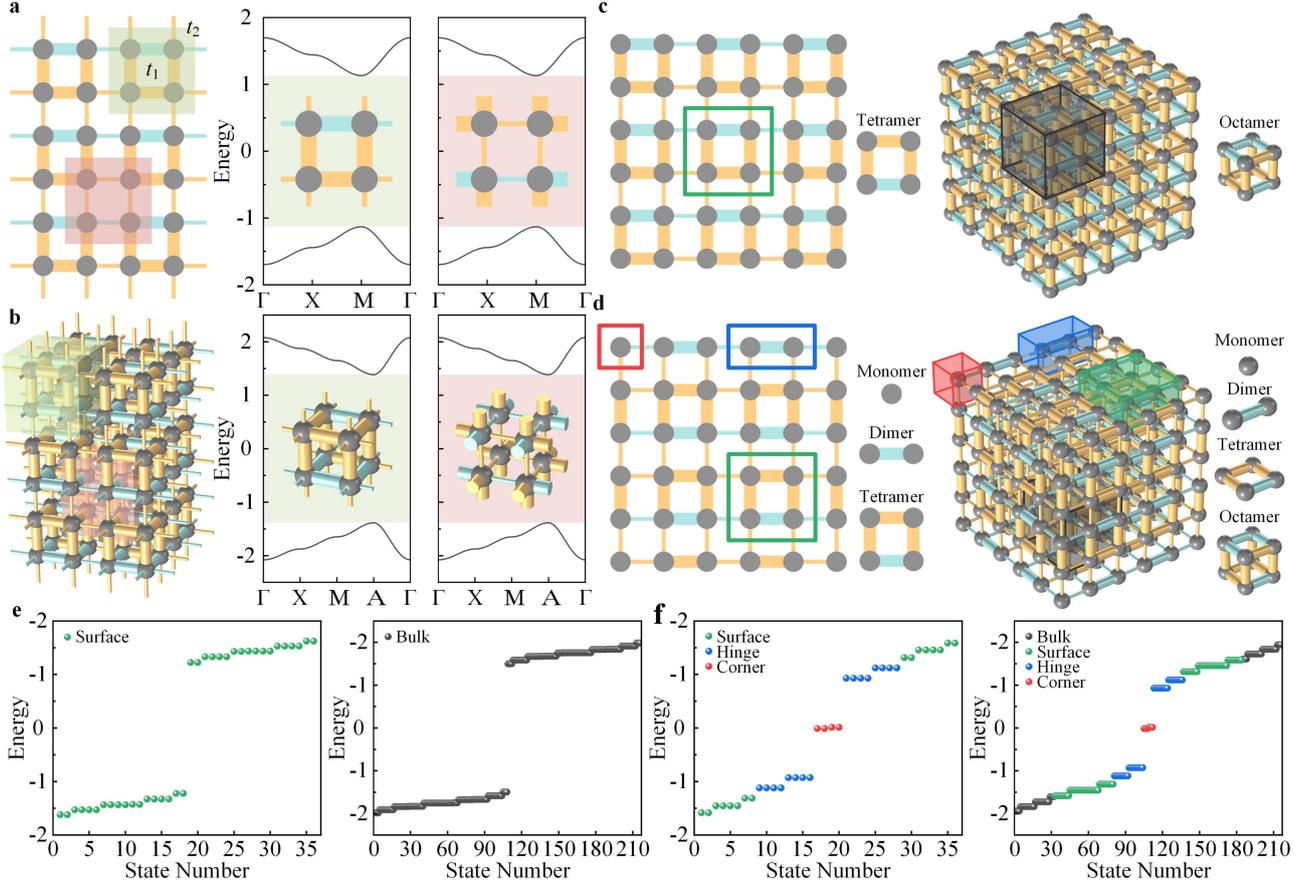

Fig. 1 | 2D and 3D tight-binding models. **a** and **b**. Illustrations of the simple 2D and 3D infinite lattice models with strong intra-unit-cell couplings $t_1 = \pm 1$ and weak inter-unit-cell couplings $t_2 = \pm 0.2$ (left). The band structures of trivial (green area) and nontrivial (red area) unit cells (right). **c**. Illustrations of the trivial lattice structures consisting of tetramers (for the 2D one) and octamers (for the 3D one). **d**. Illustrations of the nontrivial lattice structures consisting of the monomers, dimers and tetramers (for the 2D one) and the monomers, dimers, tetramers and octamers (for the 3D one). **e** and **f**. The eigenvalue spectra of the trivial and nontrivial lattice structures in c and d.

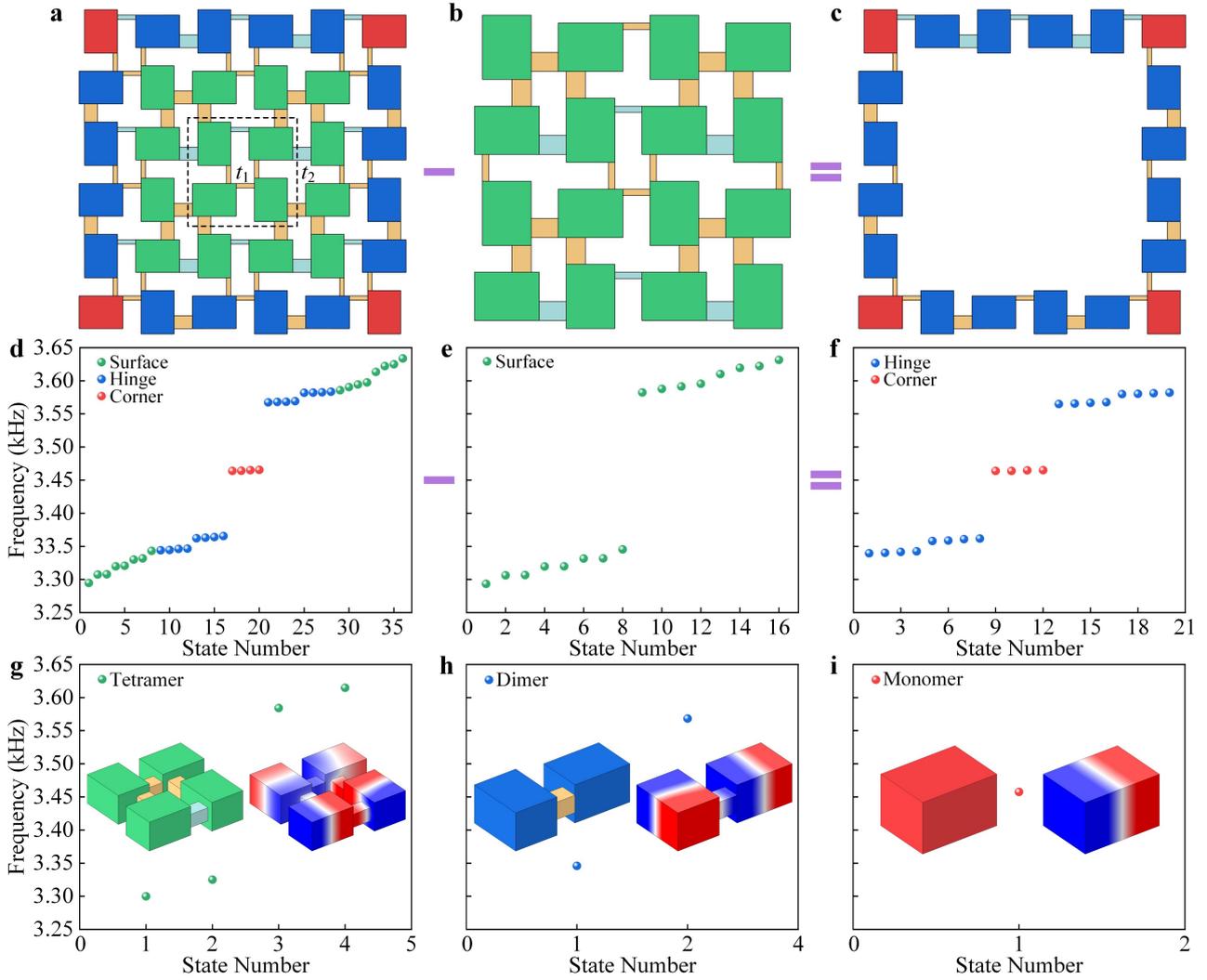

Fig. 2 | Simulated acoustic eigenvalue spectra of the 2D acoustic structure and the 2D hollow acoustic structure. **a**, **b,** and **c**. Illustrations of the conventional 2D acoustic structure, the deleted surface, and the 2D hollow acoustic structure without the surface. **d, e** and **f**. The simulated acoustic eigenvalue spectra of the conventional 2D acoustic structure, the deleted surface, and the 2D hollow acoustic structure. The surface, hinge and corner states are marked by green, blue, red circles. **g, h** and **i**. The eigenstates of the tetramer, the dimer and the monomer. Their eigenmodes are inserted in **g, h** and **i**.

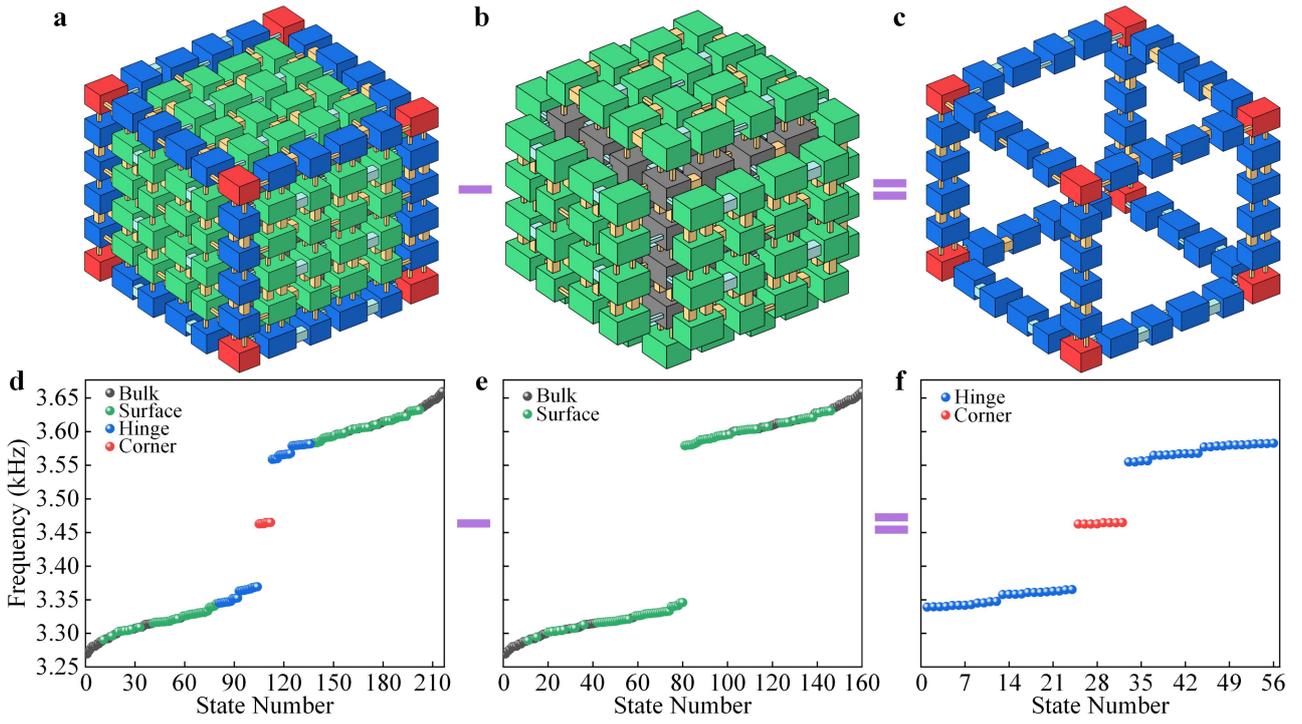

Fig. 3 | Simulated acoustic eigenvalue spectra of the 3D acoustic structure and the 3D hollow acoustic structure. **a**, **b** and **c**. Illustrations of the conventional 3D acoustic structure, the deleted bulk and surfaces, and the 3D hollow acoustic structure without the bulk and surfaces. **d, e** and **f**. The simulated acoustic eigenvalue spectra of the conventional 3D acoustic structure, the deleted surface and bulks, and the 3D hollow acoustic structure. The bulk, surface, hinge and corner states are marked by black, green, blue, red circles.

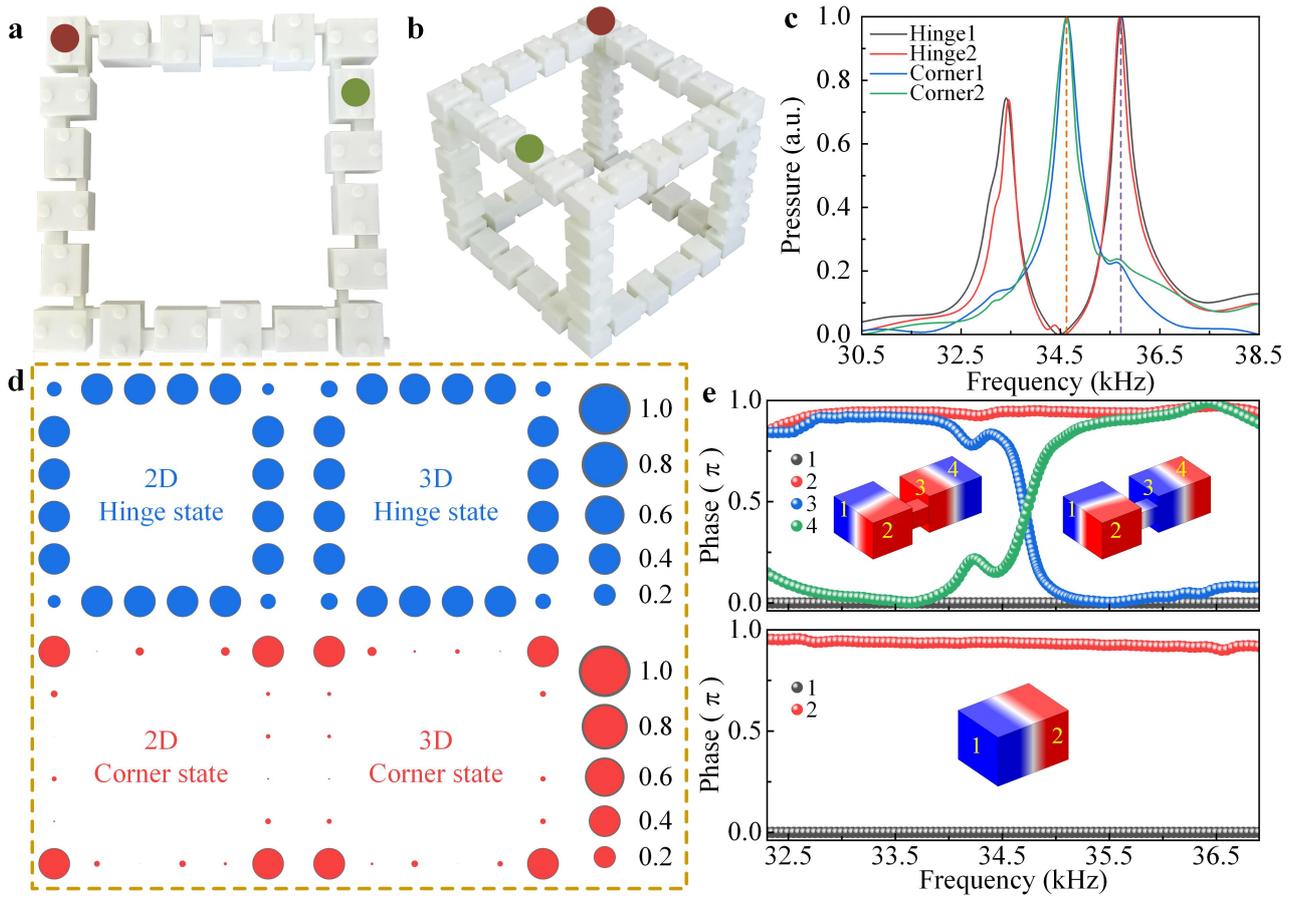

Fig. 4 | Experimental observation of HOTSs of hollow acoustic structures. **a**. Photograph of the fabricated 2D hollow acoustic structure without the surface. **b**. Photograph of the fabricated 3D hollow acoustic structure without the bulk and surfaces. **c**. Detected response spectra versus the probing frequency. Grey and red lines are the hinge response spectra of the 2D and 3D hollow acoustic structures. Blue and green lines are the corner response spectra of the 2D and 3D hollow acoustic structures. Green and brown dots (in **a** and **b**) denote the local pump-probe for the hinge and corner states. **d**. Scanned local pump-probe acoustic pressure (absolute value) distributions at the resonance peaks of the hinge and corner states. **e**. Measured acoustic phases of the hinge and corner cavities.

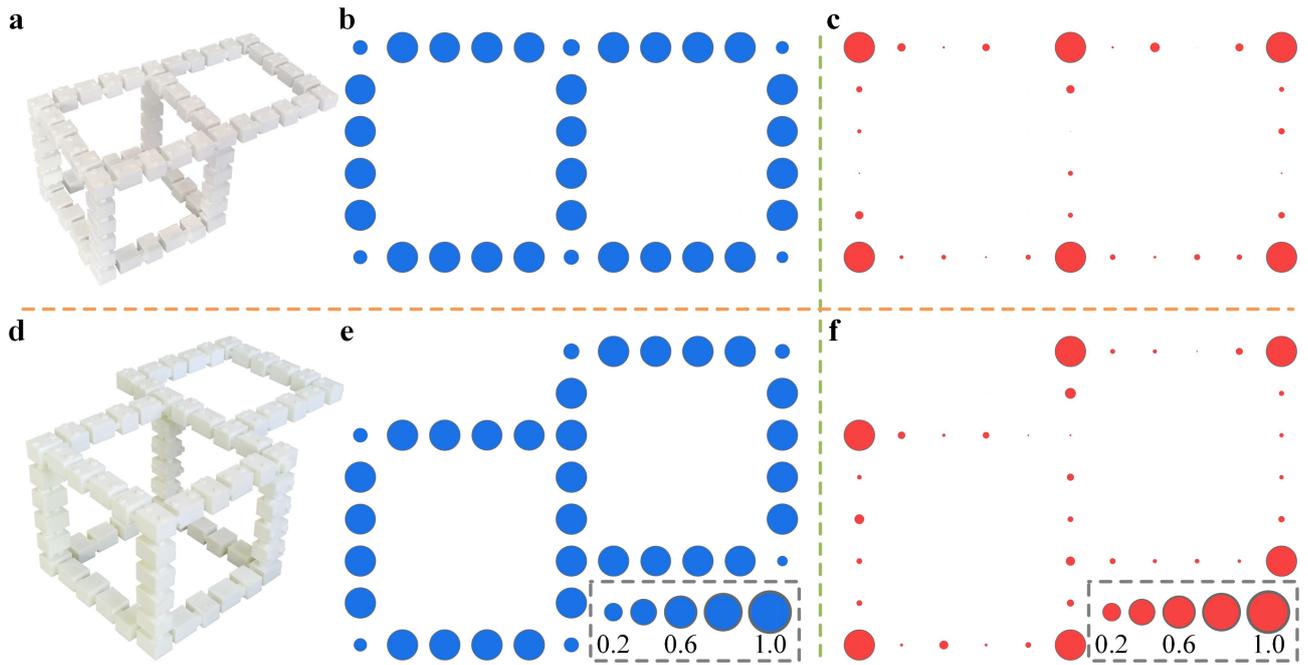

Fig. 5 | Experimental observation of HOTSs in the composite structure. **a**. Photograph of the fabricated composite structure. **b** and **c**. Scanned local pump-probe acoustic pressure (absolute value) distributions of the composite structure induced by the topological hinge and corner states. **d**. Photograph of the fabricated composite structure with the slipping dislocation. **e** and **f**. Scanned local pump-probe acoustic pressure (absolute value) distributions of the composite structure with the slipping dislocation.